\documentclass[useAMS,usenatbib]{mn2e}
\usepackage{amsmath,amssymb,epsfig,natbib,bm,psfrag,ifthen,url}
\usepackage{hyperref}
\usepackage{epsfig}
\usepackage{graphicx}
\usepackage{psfrag}
\usepackage{url}

\newcommand{\mnras}{Monthly Notices of the Royal Astronomical Society}
\newcommand{\nat}{Nature}
\newcommand{\prd}{PRD}
\newcommand{\physrep}{Physics Reports}
\newcommand{\aj}{Astronomical Journal}
\newcommand{\apj}{Astrophysical Journal}
\newcommand{\apjl}{Astrophysical Journal Letters}
\newcommand{\aap}{Astronomy and Astrophysics}

\newcommand{\apjs}{Astrophysical Journal, Supplement}

\usepackage{amsmath,amssymb}

\newcommand{\half}{\frac{1}{2}}

\newcommand{\phicen}{\Phi^{\rm cen}}
\newcommand{\phisat}{\Phi^{\rm sat}}
\newcommand{\mcen}{M_{\rm cen-cut}}
\newcommand{\msat}{M_{\rm sat}}
\newcommand{\ltot}{L_{\rm tot}}
\newcommand{\ngc}{N_{g}^{\rm cen}}
\newcommand{\ngs}{N_{g}^{\rm sat}}
\newcommand{\lmin}{L_{\rm min}}
\newcommand{\lmax}{L_{\rm max}}
\newcommand{\fredcen}{f_{\rm red,cen}}
\newcommand{\fredsat}{f_{\rm red,sat}}

\def\eprinttmp@#1arXiv:#2 [#3]#4@{
\ifthenelse{\equal{#3}{x}}{\href{http://arxiv.org/abs/#1}{#1}
}{\href{http://arxiv.org/abs/#2}{arXiv:#2} [#3]}}

\providecommand{\eprint}[1]{\eprinttmp@#1arXiv: [x]@}
\newcommand{\adsurl}[1]{\href{#1}{ADS}}

\title[The Impact of Intrinsic Alignments: Cosmological Constraints]
{The Impact of Intrinsic Alignments: Cosmological Constraints from a Joint Analysis of Cosmic Shear and Galaxy Survey Data}
\author[Donnacha Kirk, Sarah Bridle, Michael Schneider]{Donnacha Kirk$^{1}$, Sarah Bridle$^{1}$, Michael Schneider$^{2}$\\
$^{1}$Department of Physics \& Astronomy, University College London, Gower Street, London, WC1E 6BT, UK\\
$^{2}$Institute for Computational Cosmology, Department of Physics, Durham University, South Road, Durham, DH1 3LE, UK}

\begin {document}

\pagerange{\pageref{firstpage}--\pageref{lastpage}} \pubyear{2009}

\maketitle

\label{firstpage}

\begin{abstract}
Constraints on cosmology from recent cosmic shear observations are becoming increasingly sophisticated in their treatment of potential systematic effects. Here we present cosmological constraints which include modelling of intrinsic alignments. We demonstrate how the results are changed for three different intrinsic alignment models, and for two different models of the cosmic shear galaxy population. We find that intrinsic alignments can either reduce or increase measurements of the fluctuation amplitude parameter $\sigma_8$ depending on these decisions, and depending on the cosmic shear survey properties.
This is due to the interplay between the two types of intrinsic alignment, II and GI.
It has been shown that future surveys must make a careful treatment of intrinsic alignments to avoid significant biases, and that simultaneous constraints from shear-shear and shear-position correlation functions can mitigate the effects. For the first time we here combine constraints from cosmic shear surveys (shear-shear correlations) with those from ``GI'' intrinsic alignment data sets (shear-position correlations). We produce updated constraints on cosmology marginalised over two free parameters in the halo model for intrinsic alignments.
We find that the additional freedom is well compensated by the additional information, in that the constraints are very similar indeed to those obtained when intrinsic alignments are ignored, both in terms of best fit values and uncertainties.
\end{abstract}


\section{Introduction}

Weak gravitational lensing is one of the most promising observational tools available to cosmologists
for studying the recent accelerated expansion of the Universe. Images of distant galaxies appear distorted due to the bending of light as it passes through the gravitational potential of the intervening matter. The image distortion can be described as a shearing of the original galaxy shape. This ``cosmic shear'' can be exploited to study the matter distribution in the Universe and the growth of structure. Cosmic shear has the potential to be one of the most powerful probes both of dark matter and of dark energy~\citep{esoesa,detf} which together are thought to make up about $95\%$ of the energy budget of the Universe~\citep[e.g.][]{spergelea06}.


For any individual galaxy it is impossible to separate the small cosmic shear distortion from the intrinsic ellipticity of the galaxy shape. However, the light from physically close galaxies will follow a similar trajectory to the observer, passing through spacetime curved by the same gravitational fields.
Therefore these galaxies will acquire the same cosmic shear distortion. If it is assumed that galaxy shape and orientation are randomly assigned across the sky then the cosmic shear signal can be retrieved by averaging the ellipticities of a number of galaxies close on the sky.

In practice this assumption of randomly distributed galaxy shapes is unrealistic. Spatially localised galaxies are expected to have formed within the same large-scale gravitational field, which is likely to cause an alignment in their intrinsic ellipticities \citep{heavensp88,catelankb01,jing2002,aubert_pichon_colombi_2004}. See \cite{schaefer_2008_review} for a recent review. This intrinsic alignment (IA) effect will appear as a systematic error in estimates of cosmological parameters extracted from cosmic shear data, unless accurately taken into account.

Two types of IA affect the measured cosmic shear signal. Physically close galaxies form in the same large-scale gravitational potential so share a preferred ellipticity orientation. This is the Intrinsic-Intrinsic (II) correlation.
Background 
galaxies will be lensed by foreground gravitational potentials which govern the orientation of foreground galaxies. This causes an anti-correlation between the foreground/background galaxies known as the Gravitational-Intrinsic (GI) correlation.

Broadly, there are two approaches to dealing with IAs. The first, known as ``nulling'', places steps in the weak lensing pipeline which remove the IA signal. The effect of II correlations can be removed by downweighting physically close galaxy pairs \citep{heymansbhmtw04,kings03,heymansh03,takadaw04}. The GI term is more problematic as it affects all pairs of galaxies which are not physically close. In principle a particular linear combination of tomographic shear power spectra can be used to remove the GI signal if the redshifts of the galaxies are known~\citep{joachimi_schneider_2008,joachimi_schneider_2009}. The other approach tries to model the IA signal we would expect for a particular survey. These IA contributions can then be incorporated into the predictions for the measured shear signal and any free parameters in the IA model can be varied and marginalised over
\citep{detf,bridleandking,bernstein_2008,joachimi_bridle_2009}.
This is the approach we investigate. To carry this out it is necessary to make physically motivated models for IAs. Ignoring IAs completely 
will bias estimates of cosmological parameters~\citep{hiratas04,mandelbaumhisb06,hirataea07,bridleandking}.
In addition, other cosmological data sets, such as galaxy surveys, can be used to gain empirical knowledge about the
IA effect, as well as its variation with luminosity, colour and any other variables.

Some of the modelling approaches make simultaneous use of the galaxy shear - galaxy shear correlation function and the galaxy position - galaxy shear correlation function, as measured from the same imaging survey~\citep{detf,bernstein_2008,joachimi_bridle_2009}. Where ``shear'' has contributions from both gravitational and intrinsic shear.
This was first identified as an important additional statistic for learning about galaxy formation by \cite{hu_jain_2004} and was recently suggested as a powerful tool for removing IAs by~\cite{zhang09}, which was demonstrated in~\cite{joachimi_bridle_2009}.

Cosmic shear was first detected observationally a decade ago~\citep{baconre00,vanwaerbekeea00,wittmanea00,Kaiser:2000if} and the strength of 
the signal as a function of galaxy separation has since been measured by many teams, most recently by
\cite{fuea08_mnras} and~\cite{Massey:2007wb} \citep[see also][]{schrabbackea09}.
A compilation of recent cosmic shear results including a homogeneous additional treatment of systematics was made public by~\cite{benjaminea07}. The use of photometric redshift information is now starting to allow the signal to be evaluated as a function of galaxy redshift~\citep{massey_growth2007_mnras,schrabbackea09}.
The cosmological constraints in these recent papers 
are calculated assuming there are no IAs.

However, if IAs are non-negligible then ignoring them will lead to a bias on cosmological parameters. The bias in $\sigma_8$ was estimated to be between 1 and 20 per cent for a CFHTLS-like survey by \citet{mandelbaumea06} and \citet{hirataea07}. \citet{schneiderb09} and \citet{mandelbaumea09} obtained a similar result in an approximate Fisher matrix prediction. \citet{bridleandking} showed that for future surveys the equation of state parameter $w$ may be biased by order unity if IAs are significant and ignored.

\citet{heymansbhmtw04} used COMBO-17 measurements of the II signal from shear-shear correlations between galaxies close in redshift jointly with cosmic shear RCS and VIRMOS-DESCART shear-shear correlation data to constrain a simple intrinsic alignment model amplitude simultaneously with cosmological parameters. They marginalised over possible amplitudes of the II signal and found the fitted value of $\sigma_8$ was reduced by 0.03 relative to the value found when intrinsic alignments were ignored.
\citet{fuea08_mnras} estimated the amplitude of the GI signal marginalised over a range of $\sigma_8$, $\Omega_m$ values allowed by the cosmic shear data. They found it to be consistent with zero, and this conclusion held when various different scale ranges were used for each of the fits for cosmology or for the IA amplitude parameter. They concluded that they found no evidence for a non-zero GI signal.
\citet{schrabbackea09} took significant steps to reduce the contamination of their cosmic shear constraints by IAs by excluding the auto-correlations for the 5 narrowest redshift bins and excluding luminous red galaxies (LRGs) from their analysis. They also showed results using only the autocorrelations, which were similar to those excluding the autocorrelations results, leading them to conclude that intrinsic alignments are not a significant contaminant.

Many current and future surveys are planned with cosmic shear as a major design driver, in particular from the ground
the Canada-France-Hawaii Telescope Legacy Survey 
(CFHTLS)\footnote{\url{http://www.cfht.hawaii.edu/Science/CFHLS/}}, the KIlo-Degree Survey (KIDS),
the 
Panoramic Survey Telescope and Rapid Response System (Pan-STARRS)\footnote{\url{http://pan-starrs.ifa.hawaii.edu}}
surveys, 
the Dark Energy Survey (DES)\footnote{\url{http://www.darkenergysurvey.org}}, the Large Synoptic Survey Telescope (LSST)\footnote{\url{http://www.lsst.org}}, and space missions Euclid\footnote{\url{http://sci.esa.int/euclid}} and the Joint Dark Energy Mission (JDEM)\footnote{\url{http://jdem.gsfc.nasa.gov}}.

The majority of observational constraints on galaxy IAs have been carried out at low redshift~\citep{brownthd02,heymans04,mandelbaumhisb06} but these have recently been extended towards the redshifts relevant to cosmic shear surveys~\citep{hirataea07,mandelbaumea09}.
These studies have used spectroscopic galaxy surveys to calculate the galaxy position - galaxy shear correlation function and/or the galaxy shear - galaxy shear correlation function.
Constraints calculated from these correlation functions have been fitted with simple models for IAs using a fixed cosmological model.

In this paper we calculate constraints on cosmological parameters from cosmic shear data, taking into account the likely contamination from IAs for a range of IA models.
We simultaneously fit an IA model and cosmological parameters to IA correlation functions.
We then perform a joint analysis of cosmic shear and IA data, varying both cosmology and parameters within the IA model.

The paper is organised as follows: Section \ref{sec:CSandIA}
reviews the cosmic shear formalism and some basic models of IAs; Section \ref{sec:shear_shear} presents constraints from the shear-shear correlation data; Section \ref{sec:IAsplusWGL} introduces an
IA parameterisation and constraints from shear-position data; Section \ref{sec:joint} uses shear-shear and shear-position data to produce joint constraints on IA and cosmological parameters and we conclude in Section \ref{sec:conclusions}.

Throughout we assume a flat, $\rm \Lambda CDM$ cosmology. When we refer to fiducial values for cosmological parameters
we set the present day amplitude of linear mass fluctuations $\sigma_{\rm 8}$= 0.751, the baryon density $\Omega_{\rm b}$= 0.05, the matter density $\Omega_{\rm m}$= 0.3, and the Hubble parameter defined by $H_{\rm 0} = 100 h $ km s$^{-1}$ Mpc$^{-1}$ as $h$= 0.7. We assume a Harrison-Zel'dovich primordial power spectrum slope $n_s=1$.

\section{Cosmic Shear and intrinsic alignments}
\label{sec:CSandIA}

In this section we review the cosmic shear formalism before describing the origin of IAs and their effect on the measured cosmic shear signal. We then summarise three basic approaches to modelling IAs, beginning with the linear alignment model, then its non-linear extension and finally the halo model of \citet{schneiderb09}.

\subsection{Cosmic Shear}
Cosmic shear probes the expansion history and the growth of structure in the Universe using weak gravitational lensing. The effect is rather small $(\sim1\%)$ for most galaxies especially when compared to the intrinsic ellipticity of a single galaxy $(\sim 20\%)$. Therefore it is necessary to average over many galaxies to detect a coherent signal.
A convenient statistic is the Fourier transform of the two-point shear correlation function ($\xi_r$) between pairs of galaxies, the cosmic shear power spectrum $C^{GG}_{l}$. It is well approximated by projecting the three-dimensional matter power spectrum $P_{\delta}(k;\chi)$ at a redshift corresponding to a line-of-sight comoving distance $\chi$, weighted by the lensing kernel, $W(\chi)$
\begin{equation}
C^{GG}_{l} = \int_{0}^{\chi_H}{\rm{d}}\chi \frac{W(\chi)W(\chi)}{\chi^2} P_{\delta}\bigg(k = \frac{l}{\chi};\chi\bigg)
\end{equation}
where in a flat universe
\begin{equation}
W(\chi) = \frac{3}{2}
\Omega_{\rm m}
H^{2}_{0}a^{-1}(\chi)\chi \int_{\chi}^{\chi_H}\rm{d}\chi_s\, n(\chi_s)\frac{\chi_s - \chi}{\chi_s}
\end{equation}
and $a$ is the scale factor, $\chi_H$ is the comoving distance to the Hubble horizon, and the function $n(\chi_s)$ is the selection function of source galaxies per unit comoving distance, normalized as $\int d\chi\, n(\chi) = 1$.

The two-point shear correlation function is determined by the cosmic shear power spectrum via
\begin{equation}
\xi_E(\theta) = \frac{1}{2\pi}\int_{0}^{\infty} {\rm d}l\, l\, C_l J_0(l\theta),
\label{eq:Cl_to_corrfn}
\end{equation}
where $J_0$ is the zeroth order Bessel function of the first kind.

\subsection{Intrinsic Alignments}
\label{sec:IA_basics}
If galaxies were randomly oriented on the sky then the cosmic shear signal would be relatively easy to observe because the random intrinsic contributions to the observed ellipticities would average away to zero. However this is not expected to be the case.
Galaxies form within large-scale gravitational potentials, producing some ``intrinsic alignment" between galaxy shear. There are two ways in which IAs contribute to the observed shear power spectrum:

The II correlation: Intrinsic-Intrinsic galaxy alignments occur because physically close galaxies form in the same tidal field, causing an alignment of their halos or angular momentum vectors. As a result the galaxies point in the same direction. This alignment produces an increase in the measured shear power spectrum.

The GI correlation: Gravitational-Intrinsic alignments are a cross term between intrinsic ellipticity and cosmic shear. The intrinsic ellipticity of a galaxy is aligned with the density field in which it forms and this field in turn contributes to the lensing distortion of more distant galaxies. This double role causes an anti-correlation between galaxy ellipticities, because the closer galaxy may be expected to point towards the local overdensity while the distant galaxy is stretched tangentially around the overdensity. This leads to a suppression of the measured power spectrum.

The measured shear power spectrum $C_l$ therefore arises from genuine cosmic shear, combined with IA terms produced by the II and GI correlations
\begin{equation}
C_{l} = C^{GG}_{l} + C^{II}_{l} + C^{GI}_{l}.
\label{eq:shear_shear_Cl}
\end{equation}

The strength of the II term depends strongly on the depth of the survey.
A deep survey
may contain 
galaxies at many different distances from the observer, so two galaxies which appear close on the sky are most likely to be far apart along the line of sight and therefore not physically close. Since the opposite is true for a shallower survey then the IA is stronger. In addition the lensing strength is smaller when all distances are reduced, so the cosmic shear contribution is small for a shallow survey.

The II and GI lensing power spectra are calculated analogously to $C^{GG}_{l}$
\begin{equation}
C^{II}_{l} = \int^{\infty}_{0} \frac{n^{2}(\chi)}{\chi^2} P^{EE}_{\bar{\gamma}^I}(k,\chi)\rm{d}\chi
\label{eq:C_II_fn_of_PEE}
\end{equation}
\begin{equation}
C^{GI}_{l} = \int^{\infty}_{0} \frac{2W(\chi)n(\chi)}{\chi^2} P_{\delta,\bar{\gamma}^I}(k,\chi)\rm{d}\chi
\label{eq:C_GI_fn_of_PdgI}
\end{equation}
where
$k=l/\chi$,
$P^{EE}_{\bar{\gamma}^I}(k,\chi)$ and $P_{\delta,\bar{\gamma}^I}(k,\chi)$ are 
the intrinsic-intrinsic and gravitational-intrinsic power spectra respectively.

In this paper we illustrate the effect of IAs on cosmological constraints using data from the 100 Square Degree Weak Lensing
Survey~\citep{benjaminea07}.
We compare constraints from four different approaches to IAs:
\begin{enumerate}
  \item Ignoring IA effects
  \item The linear alignment model of IAs
  \item The non-linear
  alignment (NLA) model of IAs
  \item The Halo Model of IAs.
\end{enumerate}

IAs can be studied with probes other than the usual cosmic shear measurement of shear-shear correlation functions. Section \ref{sec:IAsplusWGL} below we use data from the Sloan Digital Sky Survey (SDSS)
 galaxy redshift survey shear-position correlation functions
\citep{hirataea07} 
to constrain the amplitude, 
scale and luminosity dependence of IAs. These results can be combined with models of IA in cosmic shear predictions to produce less biased constraints on cosmological parameters. Details of the IA 
parameterisations
are given below.

\subsubsection{Linear Alignment Model}
A popular simple method for modelling IAs is known as the Linear Alignment
(LA) 
model because it assumes that the intrinsic ellipticity of 
galaxies is proportional to the curvature of the primordial large scale potential \citep{catelankb01,hiratas04}.
This is thought to be most relevant for elliptical galaxies. 

In the linear alignment model the II and GI power spectra are found to first order to be
\begin{equation}
\label{eq:P_II}
P^{EE}_{\bar{\gamma}^I}(k) = \frac{C_{1}^{2}\bar{\rho}^{2}}{\bar{D}^{2}} P_{\delta}^{lin}(k)
\end{equation}
\begin{equation}
\label{eq:P_GI}
P_{\delta,\bar{\gamma}^I}(k) = -\frac{C_{1}\bar{\rho}}{\bar{D}} P_{\delta}^{lin}(k)
\end{equation}
where $P_{\delta}^{lin}(k)$ is the linear theory matter power spectrum, $C_1$ is a normalization constant, $\bar{D}(z) = (1+z)D(z)$ where $D(z)$ is the growth factor normalized to unity at the present day, and $\bar{\rho}(z)$ is the mean matter density of the Universe as a function of redshift
\citep{hiratas04}.

\subsubsection{Non-Linear Alignment Model}
This linear alignment model is likely to hold on large scales but it takes no account of non-linear growth of structure. One somewhat ad-hoc method that has been employed ~\citep{bridleandking} to rectify this deficiency is to substitute the non-linear matter power spectrum $P_{\delta}^{nl}(k)$, instead of the linear theory matter power spectrum $P_{\delta}^{lin}(k)$, into Eq.~\ref{eq:P_II} and Eq.~\ref{eq:P_GI}. In this work we use the
\cite{smithea03} 
prescription for the non-linear matter power spectrum. We call this approach the
non-linear
alignment (NLA) model.

\subsubsection{Halo Model}
The halo model of galaxy clustering \citep{scherrerb91,scoccimarroshj01} is a simple but effective predictor for galaxy clustering statistics. To first order in this model the Universe
consists of dark matter halos, distributed according to linear theory. Each halo is described by a mass, drawn from a mass function, and a density profile. The forms of these functions are generally taken from averages of n-body simulations \citep[see e.g.][for a review]{coorays02}.

Applying the halo model to IAs provides a physically motivated way to model small-scale features of the IA signal.
A simple example was provided in~\citet{schneiderb09} which we describe and use here. Each dark matter halo is taken to have a single central galaxy, positioned exactly at the centre of the spherical halo mass distribution. These central galaxies are assumed to have ellipticities determined by the large-scale density perturbations according to the linear alignment model. The satellite galaxy number density follows the density profile of the halo and their ellipticities are defined according to a distribution in the angle between the satellite major axis and the three-dimensional radius vector of the parent halo.
Shear two-point correlation functions then consist of two components: a ``one-halo term" arising from pairs of galaxies within the same halo, and a ``two-halo term" due to pairs in different halos.

For IAs the one-halo term comes from the radial alignment of the satellite galaxies and the two-halo term is dominated by the correlation of the central galaxies in each halo. Within this model one- and two-halo terms appear in both the II and GI IA power spectra
\begin{equation}
\label{eq:P_II_halo}
   P^{EE}_{\tilde{\gamma}^I} = P^{EE,1h}_{\tilde{\gamma}^I} + P^{EE,2h}_{\tilde{\gamma}^I}
   \end{equation}
   \begin{equation}
\label{eq:P_GI_halo}
   P_{\delta,\tilde{\gamma}^I} = P^{1h}_{\delta,\tilde{\gamma}^I} + P^{2h}_{\delta,\tilde{\gamma}^I}.
\end{equation}

In the case of the II power spectrum, the two-halo term can be further broken up into terms due to the correlation of satellites with satellites, satellites with centrals and centrals with centrals, and, in the GI case, the two-halo power spectrum can be separated into a central and a satellite term. However, it
was found~\citep{schneiderb09}
that the two halo terms are dominated by the central with central (II) and central (GI) terms, allowing us to ignore two-halo terms involving satellites.
This greatly simplifies the equations, allowing us to write down the two-halo terms
\begin{equation}
   P^{EE,2h}_{\tilde{\gamma}^I} = \frac{C^2_{1}\bar{\rho}}{\bar{D}}P_{\delta}^{lin}(k)
\label{eq:PEE2h}
   \end{equation}
   \begin{equation}
   P^{2h}_{\delta,\tilde{\gamma}^I} = -\frac{C_{1}\bar{\rho}}{\bar{D}}P_{\delta}^{lin}(k)
\label{eq:PdgI2h}
\end{equation}
i.e. the II and GI two-halo terms are just the same as the full linear alignment model of IAs.


\cite{schneiderb09}
provide fitting formulae for the one-halo power spectra
\begin{equation}
P^{EE,1h}_{\tilde{\gamma}^{I}}(k) = - \bar{\gamma}^{2}_{\rm scale} \frac{(k/p_1)^4}{1 + (k/p_2)^{p_3}}
\label{eq:PEE1h}
\end{equation}
\begin{equation}
P^{1h}_{\delta,\tilde{\gamma}^{I}}(k) = - \bar{\gamma}_{\rm scale} \frac{(k/p_1)^2}{1 + (k/p_2)^{p_3}}
\label{eq:PdgI1h}
\end{equation}
where $\tilde{\gamma}_{scale}$ is the
degree of alignment and determines the 
amplitude of the 1h power spectra,
that could be
set by comparison with simulations, and
for the functions $p_{i}(z)$ (for $i$ = 1,2,3) we use the fitting formulae
provided in~\cite{schneiderb09}, that are based on a halo model with redshift dependence determined by the Sheth-Tormen~\citep{shethtormen} model for the halo mass function and bias as well as the linear growth function in
their 
fiducial cosmology.

The IA power spectra $P^{EE}_{\bar{\gamma}^I}$ and $P_{\delta,\bar{\gamma}^I}$ enter both the shear-shear and shear-position calculations. In principle they depend on cosmology through the contents of the one- and two-halo terms, given in equations
\ref{eq:PEE2h} to \ref{eq:PdgI1h}.
The two-halo term contains the linear theory matter power spectrum and the growth rate, both of which depend on the values of cosmological parameters. The one-halo term is derived from a consideration of the number halos as a function of halo mass, and a match to the redshift distribution of the survey in question. It was shown in~\citet{schneiderb09} that the dependence on the survey redshift distribution is weak and thus we neglect it here. The dependence of both terms on cosmology could be included, and this would make sense if we had complete faith in the models. We have the alternative of considering the IA power spectra for our fiducial model as a template representing a reasonable but poorly understood guess at the contribution from IAs.
For this work we use the one-halo term as given in~\citet{schneiderb09} and ignore any cosmology dependence within it.
We choose to fix $\sigma_8$ within the two-halo calculation to the fiducial value.
Therefore 
the relative amplitude of the one- and two-halo terms stay roughly constant and
the overall amplitude of the sum is varied (see below). 
In this paper we choose to vary the matter density inside the linear theory matter power spectra used in the two-halo terms in equations \ref{eq:PEE2h} and \ref{eq:PdgI2h} and in the $\Omega_{\rm m}$ contribution to the normalised growth in the denominator of these equations.

To include the colour and luminosity dependence
we generalise our halo model equations~\ref{eq:P_II_halo} and~\ref{eq:P_GI_halo}
\begin{eqnarray}
P^{EE}_{\tilde{\gamma}^I}
&=&
\left(P^{EE,1h}_{\tilde{\gamma}^I} + P^{EE,2h}_{\tilde{\gamma}^I} \right)
\left[ A \left(\frac{L}{L_0}\right)^{\beta}f_r \right]^2
\label{eq:P_II_halo_general}
\\
P_{\delta,\tilde{\gamma}^I}
&=&
\left(P^{1h}_{\delta,\tilde{\gamma}^I} + P^{2h}_{\delta,\tilde{\gamma}^I} \right)
\left[ A \left(\frac{L}{L_0}\right)^{\beta}f_r \right]
\label{eq:P_GI_halo_general}
\end{eqnarray}
which scales the amplitude of both one- and two-halo terms by the same factor, given in square brackets. The factor in square brackets is squared in the first equation and not in the second equation, to mimic a simultaneous modulation of the one-halo scaling parameter $\gamma_{\rm scale}$ and the two-halo amplitude parameter $C_1$.
Throughout we retain the fiducial values of $\gamma_{\rm scale} = 0.21$ following ~\citet{schneiderb09} and $C_1 = 5\times10^{-14} (h^2 M_{\odot} {\rm Mpc}^{-3})^{-1}$ following~\citet{bridleandking}.

${L}/{L_0}$ is the normalised luminosity of the data bin and $f_r$ is the fraction of red galaxies in the data bin.
This power law in luminosity is equivalent to that used in the power law fits of~\cite{hirataea07}.
The motivation for multiplication by the red fraction $f_r$ is that we assume only red galaxies have IAs. These equations could be generalised to have different IA amplitudes for red and blue galaxies by adding a term proportional to the blue fraction $(1-f_r)$ with a different variable amplitude parameter. However, we defer such modelling to future work. A and $\beta$ are free parameters. Note that the above equations reduce to the basic halo model for $A = f_r = 1$, $\beta = 0$.

\section{Shear-shear correlations}
\label{sec:shear_shear}
In this Section we summarise the cosmic shear data we use and compare it with predicted correlation functions with
the 
various IA models. We then
make a first calculation of 
the impact of the models on constraints on the amplitude of
matter clustering and the matter density of the Universe.

\begin{figure}
\center
\epsfig{file=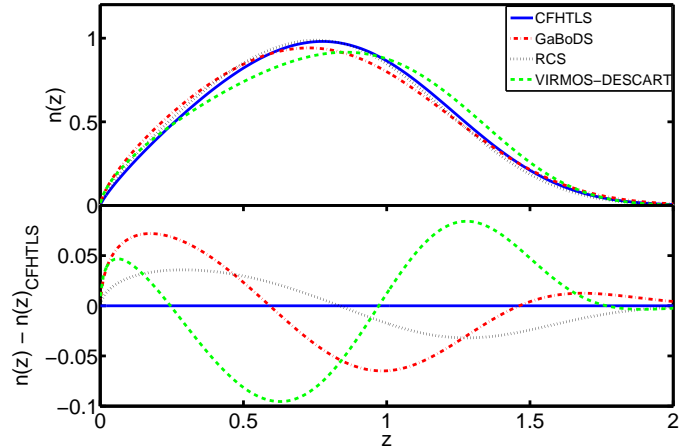,height=6cm,angle=0}
\caption{The upper panel shows galaxy redshift distribution, n(z), for the four surveys comprising the 100 deg$^2$ cosmic shear dataset. The lower panel shows the $n(z)$ difference with respect to CFHLTS, $n(z) - n(z)_{\rm CFHTLS}$. CFHTLS is the solid line, GaBoDS is dot-dashed, RCS is dotted and VIRMOS-DESCART is dashed.}
\label{fig:n(z)}
\end{figure}

\subsection{Cosmic Shear Data}

\begin{figure*}
\center
\epsfig{file=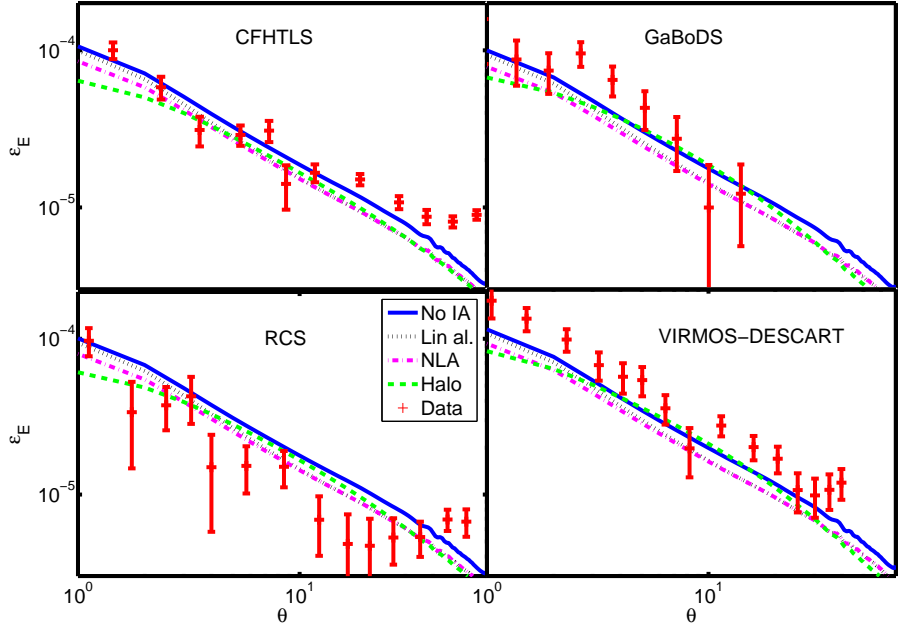,width=12cm,angle=0}
\caption{Shear correlation function, $\xi_E$, data from the four surveys comprising the 100 deg$^2$ cosmic shear dataset along with correlation functions for each of the four IA models using the fiducial cosmology and the appropriate n(z) for each survey. The correlation function without IAs is shown by the solid line, the linear alignment model as dotted line, the NLA model as dot-dashed and the halo model as the dashed line. The data, with 68$\%$ errors, appears as crosses.}
\label{fig:corrfns_100sqdeg}
\end{figure*}

The dataset used to constrain models in this paper is the 100 Square Degree weak lensing survey \citep{benjaminea07}. This combines data from the Canada-France-Hawaii Telescope Legacy
Survey (CFHTLS)-Wide, the Garching-Bonn Deep Survey (GaBoDS),
the Red-sequence Cluster Survey (RCS) and the VIRMOS-DESCART surveys.


The CFHTLS-Wide data included in this compilation covers an area of 22 deg$^2$, reaching a depth of i' = 24.5 \citep{Hoekstra:2006cs}. There is 13 deg$^2$ of data from the GaBoDS survey which uses objects which lie in the interval R $\in$ [21.5,24.5]
\citep{hetterscheidt2007}. The RCS data covers 53 deg$^2$ with a limiting magnitude of 25.2 in the $R_C$ band \citep{hoekstraea2002RCS}. The VIRMOS-DESCART data has an effective area of 8.5 deg$^2$ and a limiting magnitude of $I_{AB} = 24.5$ \citep{lefevreea04}.

Throughout this paper we use the redshift distributions for each data set as given by Eq. 9 of~\citet{benjaminea07} using the parameters given in the upper section of their Table 2 for the high confidence regime \citep[fitted to photometric redshifts in the range $0.2<z<1.5$ from][]{ilbertea06}.
\citet{benjaminea07}  
took into account uncertainties in the photometric redshift distributions by sampling from tuples of these parameter values and selecting the best fit values for each of the three parameters.
For computational practicality we here ignore uncertainties in the photometric redshift distributions. We therefore next describe in more detail the
relative 
differences between the redshift distributions of the different surveys, ignoring the
uncertainties.

Since the impact of IAs on each dataset hinges so much on the redshift distribution we show the redshift distributions and the difference in distribution, with respect to CFHTLS, for each dataset in Fig.~\ref{fig:n(z)}. We see that VIRMOS-DESCART and GaBoDS have the greatest number of low redshift galaxies and VIRMOS-DESCART shows a signifigant
excess 
at $z\sim1.4$.

\subsection{Correlation Functions Including Intrinsic Alignments}
\label{sec:IAsonWGL}

Fig. \ref{fig:corrfns_100sqdeg} shows the correlation functions of the four constituent surveys of the 100 deg$^2$ dataset.
We overlay the predicted correlation functions
for each of the IA models we consider.
The correlation functions are calculated for the fiducial cosmology, with no attempt to find the best fit to each survey. Note that Fig. 1 in \citet{benjaminea07} plots similar calculated correlation functions but uses a different, best-fit, $\sigma_{\rm 8}$ for each survey.

Overall the GI contribution dominates the IA signal and has the effect of reducing the shear-shear correlation signal; hence all the correlation functions which include an IA component are mostly lower than the equivalent correlation function without IAs (solid line). As expected, the halo model (dashed line) agrees with the LA (dotted line) and NLA (dot-dashed line) models at large
angular separation 
where the two-halo term and linear theory dominate.
The halo and
LA 
models diverge at small
angular separation 
when the additional one-halo term becomes important
and the 
shear-shear correlation amplitude is reduced due to the negative contributions from the GI term. The NLA model is intermediate between the LA and halo model at
small angular separations. 
The increase in the halo model correlation function at intermediate scales is caused by an increased II contribution on these scales.
In the halo model of IAs the galaxies align
towards the center of 
dark matter halos, and as a result are also aligned with each other.
Therefore 
the one-halo 
term for the GI signal dominates on a scale corresponding to the typical distance of galaxies from the halo center, whereas the one-halo 
 term for the II signal occurs at the typical halo diameter. This explains why the GI term dominates at small scales and the II term has a bigger impact at intermediate scales.

We now compare the predicted correlation functions for the different surveys. The only difference between the predictions comes from the different redshift distributions used.
The IA signals dominate over the cosmic shear signal at low redshift. A large number of physically close galaxies produces a large II effect.
Surveys with a large number of low redshift galaxies will have a large proportion of these physically close
strongly aligned galaxies. Fig.~\ref{fig:n(z)} shows that VIRMOS-DESCART (dashed line) and GaBoDs (dot-dashed line)
have the largest proportion of low redshift galaxies, so we expect the strong increase at intermediate scales in Fig.~\ref{fig:corrfns_100sqdeg}. Conversely CFHTLS has the lowest proportion of low-z galaxies and we see little increase in its correlation function due to an II contribution.

\subsection{Impact on $\sigma_{\rm 8}$ constraints}

\begin{figure}
\center
\epsfig{file=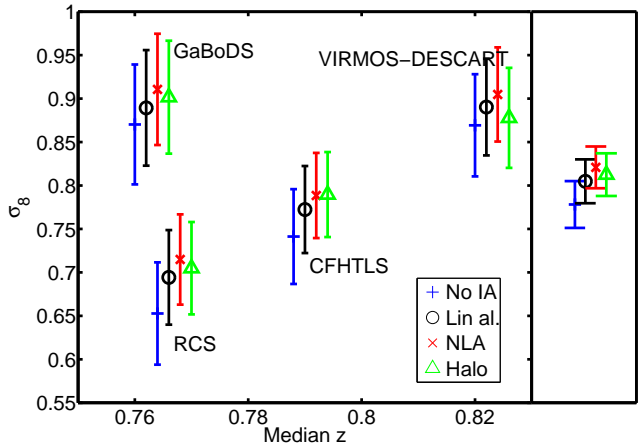,height=6cm,angle=0}
\caption{68$\%$ confidence limits on
the amplitude of fluctuations
$\sigma_{\rm 8}$ for each survey within the 100 deg$^2$ cosmic shear dataset.
From left: GaBoDS, RCS, CFHTLS, VIRMOS-DESCART, and combined. Each IA model is shown separately for
each survey, plotted at the median redshift of each survey.
From left: No IAs (horizontal crosses), linear alignment (circles), NLA model (diagonal crosses) and halo model (triangles). IA points are offset from the median redshift for clarity.
The combined constraints for the whole 100 deg$^2$ dataset are shown in the panel on the right. This analysis varies $\sigma_{\rm 8}$ leaving other parameters fixed at their fiducial values, including $\Omega_{\rm m}=0.3$.
}
\label{fig:sigma8}
\end{figure}

We are interested in the biasing of cosmological constraints when IAs are erroneously ignored.
We compare predicted and observed correlation functions and produce $\chi^2$ values at a range of 
amplitudes 
of matter clustering, parameterised by $\sigma_8$. These are used to calculate a best fit $\sigma_8$ and corresponding 68$\%$ confidence limits. This was done for each of the four cosmic shear datasets separately and the analysis was repeated using the four different IA approaches. The results are shown in Fig.~\ref{fig:sigma8}. The right-hand panel shows the 68$\%$ confidence limits using the combined 100 deg$^2$ data,
obtained by multiplying together the probabilities ${\rm Pr}(\sigma_8)\propto \exp(-\chi^2/2)$ from each survey. 
In this analysis, all other cosmological parameters are held at their fiducial values
and we fix $A=1$, $\beta=0$ and $f_R=1$. 
In all analysis quoted in this paper, we kept $\sigma_8$ constant, at its fiducial value, within the IA models themselves 
which can be scaled by an overall amplitude parameter $A$. 
As usual, the cosmic shear correlation function has a strong dependence on $\sigma_8$. 

The results show a dependence of the measured $\sigma_{\rm 8}$ on the IA model. Ignoring IAs
(horizontal/vertical crosses) 
produces results which are consistently below the results when IAs are taken into account, irrespective of the IA model used.
This behaviour is consistent with the correlation functions seen in Fig.~\ref{fig:corrfns_100sqdeg}. The dominance of the GI term has the effect of lowering the overall shear-shear correlation function prediction meaning that constraints using an IA model favour a larger value of $\sigma_{\rm 8}$ to compensate for the reduced amplitude. The effect is more pronounced for the NLA model (diagonal crosses) than for the linear alignment model (circles) because suppression of the correlation function due to the GI term is stronger in the NLA. When the halo model is used (triangles) the best fit $\sigma_{\rm 8}$ is larger than that when IAs are ignored, but the difference can be smaller than for the NLA or LA models due to the contribution from the II correlations at intermediate separations which partially cancel the GI terms.
We note that the difference between the halo model constraint and that when IAs are ignored is least for GaBoDS and VIRMOS-DESCART. This is consistent with their relatively large number of low-redshift galaxies. This produces
an 
increased II contribution at intermediate scales, lowering constrained $\sigma_{\rm 8}$ values. 
The effect is biggest for VIRMOS-DESCART because it has the greatest constraining power at these intermediate scales.

Although IAs are an important cosmic shear systematic and a clear biasing is apparent,
it is encouraging to note that the effect is not catastrophic: the best-fit values for almost all surveys and IA models lie within the 68$\%$ confidence limits of the results in which IAs are ignored.
Because the uncertainties on $\sigma_{\rm 8}$ are smaller for the joint constraints (right hand panel of Fig.~\ref{fig:sigma8}), the total biases are now similar to the 68 per cent confidence limits. We use a more sophisticated model for how IAs affect cosmic shear constraints in 
Section~\ref{sec:joint}.

\begin{figure}
\center
\epsfig{file=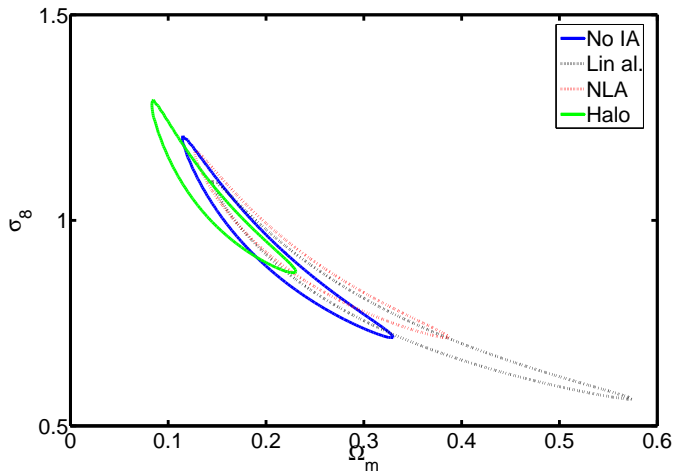,width=9cm,angle=0}
\caption{68$\%$ Confidence limits in $\sigma_8$ - $\Omega_{\rm m}$ parameter space combining the four surveys in the 100 deg$^2$ cosmic shear dataset for each IA model. Constraints where IAs are ignored in the modelling are shown by the dark, solid (blue) contour; constraints assuming the linear alignment model are shown by the dark dotted (black) contour; the NLA model by light dotted (red) and the halo model by the light solid (green) contour. IA parameters are set to $A=1$, $\beta = 0$, $f_r = 1$.
All other cosmological parameters were fixed at their fiducial values.}
\label{fig:CLs_s8Odm}
\end{figure}


\subsection{Impact on constraints on $\sigma_{\rm 8}$ and $\Omega_{\rm m}$}

We next calculate constraints from the cosmic shear data allowing both the fluctuation amplitude $\sigma_{\rm 8}$ and
the matter density $\Omega_{\rm m}$ to vary. Other cosmological parameters are held at their fiducial values. The results are shown in Fig. \ref{fig:CLs_s8Odm} as 68$\%$ confidence contours in $\sigma_{\rm 8}$-$\Omega_{\rm m}$ parameter space.
All results for the various IA models show the usual cosmic shear degeneracy between $\sigma_{\rm 8}$ and $\Omega_{\rm m}$. As expected, at a fixed $\Omega_{\rm m}=0.3$ we can read off the $\sigma_8$ constraints shown in the previous figure.
However the relative pattern between models depends on the assumed value of $\Omega_{\rm m}$ and furthermore different $\Omega_{\rm m}$ ranges are preferred by different models.

\begin{figure*}
\center
\epsfig{file=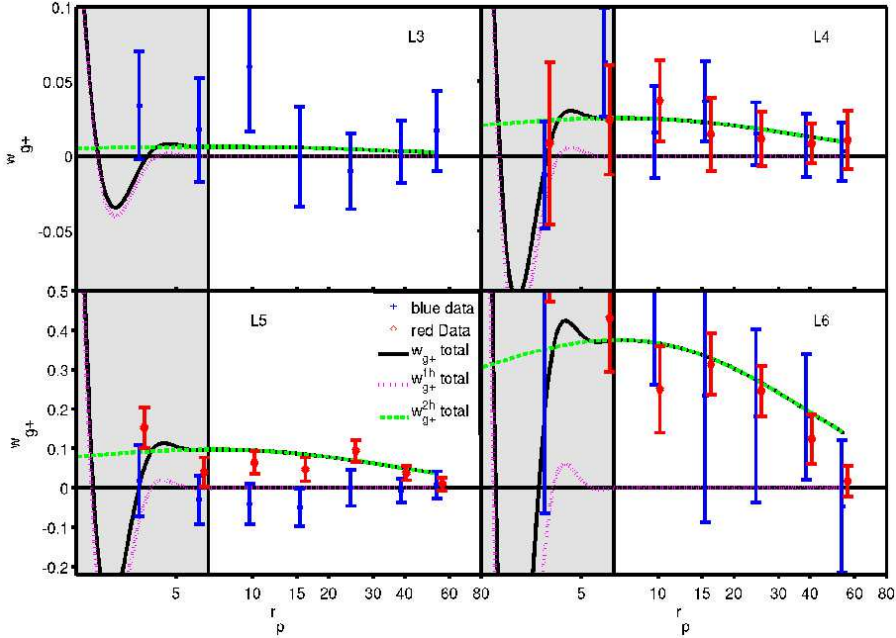,width=12cm,angle=0}
\caption{Comparison of the shear-position correlation function, $w_{g+}$, from the~\citet{hirataea07} data and the halo model prediction using fiducial cosmology, A = 1, $\beta = 1.44$, $L_0 = L_4$. Predicted correlation functions are shown for the one-halo term only (dotted), two-halo term only (dashed) and the total (solid). Blue galaxy data for each luminosity bin is plotted as crosses, red galaxies as diamonds. The shaded region shows distance scales excluded from our analysis.
}
\label{fig:corrfns_hirata}
\end{figure*}

The most interesting result is that from the halo model of IAs (light solid (green) contour), which favours the high $\sigma_{\rm 8}$ and low $\Omega_{\rm m}$ end of the degeneracy. Very roughly speaking the shear-shear correlation function data provide two pieces of information: an amplitude and a slope. The amplitude constrains a degenerate combination of $\sigma_8$ and $\Omega_{\rm m}$ 
but the slope of the correlation function partly breaks this degeneracy and determines the preferred range range of $\Omega_{\rm m}$ along the degeneracy line.
The most constraining data set is CFHTLS, that has tightest constraints on large scales. The effect of the halo model of IAs is to steepen the correlation function on these scales due to the
contribution from the II one-halo term at intermediate scales 
(Note that this one-halo 
II term does not vary with cosmological parameters in our implementation so this conclusion stands at all positions in cosmological parameter space.) An increase in $\Omega_{\rm m}$ flattens the matter power spectrum
on the scales of interest,  
thereby increasing the amount of power on small scales relative to large scales.
This means that an increase in $\Omega_{\rm m}$ leads to a more steeply decreasing correlation function. Thus to reproduce the observed data in the presence of IA halo model contributions $\Omega_{\rm m}$ must decrease, leading to the position of the halo model contours.

\section{Shear-position correlations}
\label{sec:IAsplusWGL}

In this Section we introduce a new data set which was produced with the aim of measuring the IA signal. We use the galaxy shear - galaxy position two-point function that quantifies whether the shape of one galaxy points at the position of another galaxy.
If we have a good understanding of how the galaxy positions are related to the mass distribution then this effectively constrains the IA power spectrum $P_{\delta,\tilde{\gamma}^I}$. We compute predictions for the two point function using the halo model for IAs, then 
compare it to the data to measure IA model parameters.

\subsection{Data}

We use data measured from the SDSS galaxy survey, a 5-band (ugriz) photometric survey over~$\pi$ steradians of the sky, with spectroscopic follow-up of $~10^6$ objects \citep{SDSS_mnras}.
We use the galaxy position - galaxy shear correlation functions
from the data described in  \citet{mandelbaumhisb06} and calculated in 
\citet{hirataea07}. These are calculated from the main spectroscopic sample (z = 0.05 - 0.2).

This is split by luminosity and colour
and 
each of these subsamples is used to trace the intrinsic shear field. The luminosity subsamples span L3 (one magnitude fainter than $L_*$) to L6 (two magnitudes brighter). Each of these is divided into ``red'' and ``blue'' colour subsamples, defined using an empirically determined, redshift dependent separator, $ u- r = 2.1 + 4.2z$ in the observer frame. The full sample is used to trace the density field. Ellipticity measurements were taken from \citet{mandelbaumea2005a}.

\subsection{Correlation Functions}

The correlation function is calculated by taking pairs of galaxies where one galaxy is from the shear catalogue (luminosity and colour bin) and the other is from the density catalogue (which is a composite of all the luminosity and colour bins). The ellipticity of the galaxy from the shear catalogue is calculated in a coordinate system aligned with the vector between the two galaxies in the pair. If the galaxy in the shear catalogue is aligned along the vector joining the pair then this number will be positive. If it points perpendicular to this line then the ellipticity in this coordinate system will be negative. These ellipticities are averaged over all pairs, as a function of separation on the sky $r_p$ and the separation along the line of sight $\Pi$, taking into account edge effects using a generalized Landy-Szalay~\citep{landy_szalay} estimator.
This is integrated over the range $-60 < \Pi / h^{-1} {\rm Mpc} <60$. We use the covariance matrices between bins in angular separation which were provided.


We predict the shear-position correlation function $w_{g+}$ using
\begin{equation}
w_{\delta +} = -\frac{1}{2\pi} \int P_{\delta,\tilde{\gamma}^I}(k)J_{2}(kr_p) \, k \, \rm{d}k
\label{eq:wdeltaplus}
\end{equation}
\citep[][Eq. 23]{hirataea07} which we assume is related to the galaxy correlation function by the galaxy bias $b_g$
\begin{equation}
w_{g+} = b_{g}w_{\delta+}
\label{eq:wgplus}
\end{equation}
where we use galaxy bias estimates from Table 3 of \citet{hirataea07} and we allow for the variation of galaxy bias with $\sigma_{\rm 8}$ as they describe. 
We use the generalised halo model as given in Eq.~\ref{eq:P_II_halo_general} and Eq.~\ref{eq:P_GI_halo_general} and take values for the luminosity of each SDSS shear-position correlation function bin from the minimum magnitude in each bin as given in~\cite{hirataea07} Table 1. We set $L_0$ to be the luminosity of the L4 bin. We set the red fraction $f_r$ to be unity for the red samples and zero for the blue samples, so that effectively we do not use the blue samples in the constraints. Following~\cite{hirataea07}, we use only the data points for separations greater than $4.7 h^{-1}$ Mpc where the bias can reasonably be assumed to be linear.
The overall amplitude scaling parameter $A$ and luminosity power law $\beta$ are the two free parameters we consider.

Fig.~\ref{fig:corrfns_hirata} shows the shear-position correlation function, $w_{g+}$, for each luminosity and colour bin in the~\citet{hirataea07} data (crosses for blue galaxies,
diamonds for red galaxies). 

Also shown are predictions using the halo model, using for illustration $A = f_r = 1$ and $\beta = 1.44$, the best-fit
$\beta$ value provided by the third row of the lower half of table 4 in \citet{hirataea07}.
Separate correlation functions are plotted for the one-halo contribution (dotted line), two-halo contribution (dashed line) and the total (solid line). As we would expect, the two-halo term provides the bulk of the GI signal at high separation, and falls off at small separation. An interesting feature of the one-halo correlation function is that as the separation is decreased it rises from zero, then falls to negative values, before rising sharply at low separation.
This behaviour was discussed in~\citet{schneiderb09} with reference to their Fig.~2 in their Section 2A.
The grey-shaded regions on Fig.~\ref{fig:corrfns_hirata} show the scales which have been excluded from our analysis of the IA data, following the practice of \citet{hirataea07}. We see that essentially only the two-halo term is significant in the fits to this dataset.

\subsection{Constraints on Intrinsic Alignment Parameters}

\begin{figure}
\center
\epsfig{file=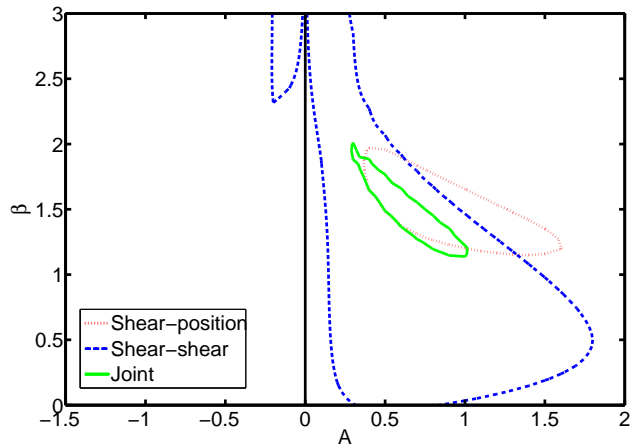,height=6cm,angle=0}
\caption{68$\%$ confidence limits on A and $\beta$, the amplitude and luminosity dependence of the IA power spectra.
Constraints are plotted from the~\citet{hirataea07} shear-position data (dotted contour), the 100 deg$^2$ shear-shear data (dashed contour) and joint constraints (solid contour).
$\sigma_{\rm 8}$ was marginalised over and all other cosmological parameters were fixed at their fiducial values.}
\label{fig:Cls_A_beta}
\end{figure}

In this section we use the shear-position correlation function data to constrain A and $\beta$, the  free parameters in the IA model. The dotted line in Fig.~\ref{fig:Cls_A_beta} shows the resulting 68$\%$ 
 confidence limits. Both shear-shear and shear-position data are marginalised over $\sigma_{\rm 8}$. 
We chose to pivot the luminosity power law at the SDSS L4 luminosity bin ($L_0$ corresponds to the magnitude of the L4 bin).
The red galaxies are either in luminosity bin L4 or higher luminosity bins,
therefore an increase in $\beta$ will
mostly 
increase the predicted IA signal relative to the data. The same is true for $A$, therefore we expect them to be anti-correlated, as seen in Fig.~\ref{fig:Cls_A_beta}.
We discuss the remaining lines on this figure in the following section, but note here that the shear-position correlation data is by far the most important dataset when constraining these two parameters alone.


Our results agree well with those of~\cite{hirataea07} who find $\beta=1.44^{+0.63}_{-0.62}$ when using all the SDSS Main data (blue and red) simultaneously, with the same minimum separation in the correlation function.
By construction the best fit $A$ is around unity, since the vast majority of the constraint comes from the two-halo term of the halo model which corresponds to the linear theory prediction estimated from~\cite{hiratas04} which was fitted to the~\cite{heymansbhmtw04} data which gave a similar amplitude as the~\cite{mandelbaumhisb06} result.

To allow a comparison between the cosmic shear data and the IA model used for the galaxy survey data values of $f_r$ and $L$ were calculated
as a function of redshift 
for each component survey of the 100 deg$^2$ dataset using a conditional luminosity function model based on \citet{cooray2006} and outlined in appendix~\ref{sec:clfmodel}.

In the previous section we computed cosmological constraints from cosmic shear data with and without consideration of IAs. The ``No IA'' results correspond to $A=0$ on this Figure, that is clearly ruled out by the SDSS data. The ``halo model'' results corresponded to $A=1$, $\beta=0$, $f_r=1$ which is also ruled out by this figure but would be appropriate if all the galaxies in the cosmic shear survey had the same luminosity and colour as the SDSS Red L4 sample. In the following Section we present a full joint analysis of both samples.

\section{Joint shear-shear and shear-position constraints}
\label{sec:joint}

In this Section we compare constraints on parameters from shear-shear and shear-position correlation functions and combine the constraints to produce limits on both IA and cosmological parameters.
To make our previous analysis of the cosmic shear data more general it would be good practice to allow
the IA free parameters A and $\beta$ to vary, then marginalise over them rather than assuming they were fixed to the basic halo model prescription. This would inevitably lead to a loss of constraining power. However, this loss can be made less severe by calculating joint constraints on IA parameters from cosmic shear and galaxy survey data. With this approach, the extra information about IAs provided by galaxy surveys is used to produce less biased cosmological constraints from cosmic shear.

\subsection{Joint Constraints on Intrinsic Alignment Parameters}

\begin{figure}
\center
\epsfig{file=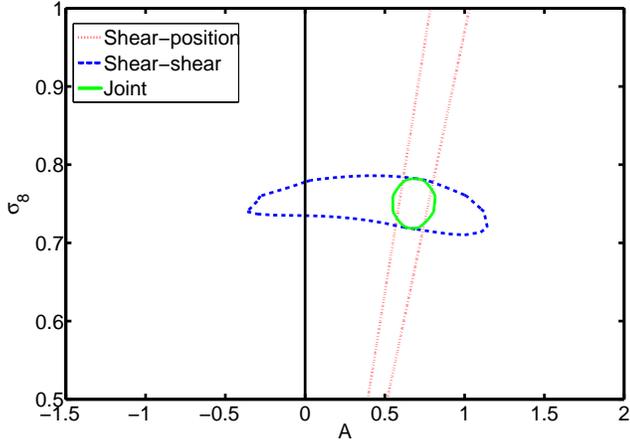,height=6cm,angle=0}
\caption{68$\%$ confidence limits on $\sigma_{\rm 8}$ and $A$, the IA power spectrum amplitude parameter. Constraints are plotted from the~\citet{hirataea07} shear-position correlation function data (dotted lines), the \citet{benjaminea07} 100 deg$^2$ shear-shear correlation data (dashed contour) and joint constraints (solid contour). The IA luminosity dependence power law slope was fixed $\beta = 1.44$, and all other cosmological parameters were fixed at their fiducial values.
} \label{fig:7}
\label{fig:Cls_s8_A}
\end{figure}

We combine constraints from the two datasets by simply adding together their $\chi^2$ values. This is acceptable given that the surveys do not overlap significantly. We calculate $\chi^2$ values from both datasets as a function of both IA parameters and cosmological parameters.
For the shear-shear dataset the IA parameters enter via the IA power spectra (Eq.~\ref{eq:P_II_halo_general} and Eq.~\ref{eq:P_GI_halo_general}) which are projected onto the sky (Eq.~\ref{eq:C_II_fn_of_PEE} and Eq.~\ref{eq:C_GI_fn_of_PdgI}) and added to the cosmic shear contribution to produce the full shear-shear power spectrum (Eq.~\ref{eq:shear_shear_Cl}) and the correlation function is calculated by Eq.~\ref{eq:Cl_to_corrfn}.
The shear-position correlation functions are computed from the shear-density correlation function (Eq.~\ref{eq:wgplus}) which is a inverse Hankel transform (Eq.~\ref{eq:wdeltaplus}) of the relevant IA power spectrum (Eq.~\ref{eq:C_GI_fn_of_PdgI}).
The bias values given in~\cite{hirataea07} are calculated from measurements of galaxy clustering (position-position correlation functions) and therefore depend on the assumed value of $\sigma_8$. We take this into account to obtain a bias values for each $\sigma_8$ value considered. 
As discussed at the end of Section~\ref{sec:IA_basics}, we fix $\sigma_8$ in the IA power spectra and allow $\Omega_m$ to vary in the linear theory matter power spectrum contribution to the two-halo term.

First we allow only $\sigma_{\rm 8}$ and $A$, the amplitude of the IA signal, to vary, with the rest of the cosmological parameters set to their fiducial values and a fixed luminosity dependence power law slope $\beta = 1.44$
\citep[the best-fit value given in][]{hirataea07}. The results are shown in Fig.~\ref{fig:Cls_s8_A} as 68$\%$ confidence contours in $\sigma_8$-$A$ parameter space. Contours are shown for the shear-position correlation functions
(nearly vertical lines) 
shear-shear correlation functions
(nearly horizontally elongated contour)
and the combined constraint
(roughly at the intersection). 
The shear-position correlation function constraints on the IA amplitude parameter $A$
for the fiducial $\sigma_8$ value 
are as expected from Fig.~\ref{fig:Cls_A_beta} for the fixed $\beta$ value, with $A$ significantly larger than zero.
In the analysis of the shear-position correlation function data $\sigma_{\rm 8}$ is held fixed in in the calculation of all our IA models. However, the variation of galaxy bias as a function of $\sigma_{\rm 8}$ produces the expected degeneracy with $A$. Increasing $\sigma_{\rm 8}$ above the fiducial value decreases galaxy bias, requiring a greater $A$ to compensate and vice-versa. 

The shear-shear correlation functions do themselves place some constraint on the IA parameter $A$, preferring a range of order unity. A negative value of $A$ would correspond to galaxies pointing in the opposite direction to that used in the standard models. The linear alignment model (two-halo term) would contain galaxies pointing perpendicular to the tidal stretching expected by the gravitational potential curvature. The one-halo picture would contain galaxies which are aligned tangentially to the center of the halo. We described earlier a rough picture of shear-shear correlation function constraints in which essentially the data measure the amplitude and slope of the correlation function. The amplitude essentially fixes a degenerate combination of $\sigma_8$ and $A$, and the constraint on $A$ must therefore come from the shape of the correlation function. For the fiducial $\Omega_{\rm m}$, Fig.~\ref{fig:Cls_s8_A} tells us that a large IA contribution to the shear-shear power spectra distorts them too much. This can be understood by examining Fig.~\ref{fig:corrfns_100sqdeg}, in which the data points fit well with the shape of the lensing-only (``No IA") predictions, whereas the halo model predictions tend to be more curved over the scales probed.

The direction of degeneracy between $A$ and $\sigma_8$ from shear-shear information alone can also be understood in terms of the shear-shear correlation function datapoints in Fig.~\ref{fig:corrfns_100sqdeg}. In general there is a balance of effects between the II and GI contributions. If we imagine a universe in which the GI term did not exist but the II term did, an increase in the IA amplitude parameter $A$ would add power to the predicted
shear-shear correlation function. To keep the predictions consistent with the data points it would be necessary to decrease $\sigma_8$ to reduce the lensing contribution. This would give a negative correlation between $A$ and $\sigma_8$, seen for larger positive $A$ values. In this unphysical universe containing only II terms, the direction of degeneracy would appear reversed for negative $A$ because $A$ appears as a squared quantity in the II terms. Indeed we do see the direction reversed in this figure. In the physical Universe the effect of the GI contribution is to break the symmetry slightly, and prefer slightly more positive $A$ values since it causes a slight cancellation in the IA effect for positive $A$ values
which fits better to the shape of the correlation functions that an addition of power. 

The great complementarity of the two datasets is most clearly illustrated in this figure. The main constraints on cosmology come from the shear-shear correlation data and the main constraints on IAs from the shear-position data. The joint constraints focus at the intersection between the two relatively degenerate constraints, as expected.

The analysis was repeated, allowing $A$, $\beta$ and $\sigma_{\rm 8}$ to vary. We marginalised over $\sigma_8$ to produce the results shown in Fig.~\ref{fig:Cls_A_beta}. All other parameters were fixed at their fiducial values.
The shear-shear correlation function data (dashed) provides much weaker constraints in the A-$\beta$ plane- 
partly due to marginalisaion over $\sigma_{\rm 8}$ in the power spectrum but primarily because the cosmic shear data is not broken into luminosity bins, so has very little power to constrain $\beta$. Even so we can distinguish the same degeneracy direction as present in the shear-position data and positive $A$ is favoured at 68$\%$ confidence. 
Most of the constraining power in the joint constraints (solid) comes from the shear-position data (dotted). 

\begin{figure}
\center
\epsfig{file=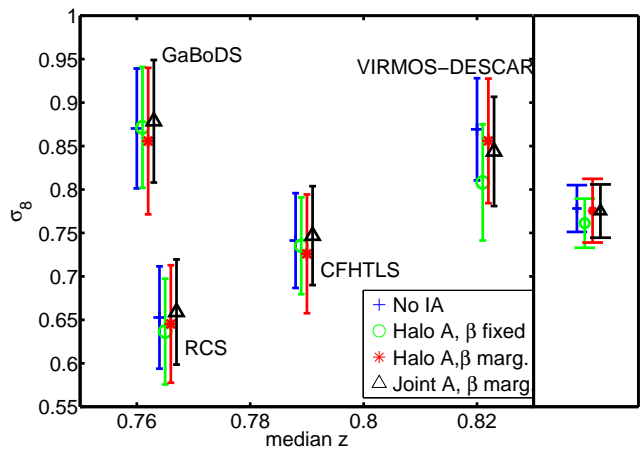,height=6cm,angle=0}
\caption{68$\%$ confidence limits on $\sigma_{\rm 8}$ for each survey within the 100 deg$^2$
shear-shear correlation function dataset (from left: GaBoDS, RCS, CFHTLS, VIRMOS-DESCART and combined). For each survey constraints shown, from left: without IAs (crosses); using the halo model with $A=1$, $\beta=1.44$ (circles); halo model, marginalised over $A$ and $\beta$ (stars); joint shear-shear and shear-position constraints, marginalised over $A$ and $\beta$ (diamonds). The right-side panel shows joint constraints from the four surveys. All other cosmological parameters were fixed at their fiducial values.}
\label{fig:s8_z_IA}
\end{figure}

\subsection{Joint Constraints on Cosmology}

Here we revisit the constraints on cosmological parameters from Section \ref{sec:IAsonWGL} showing how they change when we allow the IA parameters ($A$ and $\beta$) to vary and be marginalised over. Joint shear-shear and shear-position constraints are also presented to show constraints on cosmology when IAs are taken into account as well as possible given current models and data.

Fig.~\ref{fig:s8_z_IA} shows constraints on the amplitude of matter fluctuations $\sigma_{\rm 8}$ for each shear-shear survey separately as a function of survey median redshift, and also the joint constraint for all shear-shear surveys in the right hand panel.
The blue crosses in Fig.~\ref{fig:s8_z_IA} are the same as those in Fig.~\ref{fig:sigma8} and show constraints on $\sigma_8$ when IAs are ignored.
In Fig.~\ref{fig:sigma8} we illustrated the effect of IAs on the different IA models using a simple version of the IA power spectra in which $A=1$, $\beta=0$ and we assumed $f_r=1$. In Fig.~\ref{fig:s8_z_IA} we use only the halo model and use the values for $L$ and $f_r$ as a function of redshift as discussed in
Appendix 
\ref{sec:clfmodel}.
For the circles in  Fig.~\ref{fig:s8_z_IA} we fixed $A=1$, $\beta=1.44$, and we see that the $\sigma_8$ results using the halo model are all now lower than before.
The result of using the more realistic $L$ and $f_r$ values instead of $\beta=0$, $f_r=1$ is to lower all the IA contributions. This is to be expected since $f_r$ is always less than or equal to unity in the more realistic model, and in addition $f_r$ is particularly small at low redshifts where faint satellite galaxies dominate and the IA power spectra are largest. The biggest modification to the predicted correlation functions is the reduced GI contribution, especially on smaller scales.
Therefore the total shear-shear correlation is increased and the fitted $\sigma_8$ is reduced to compensate.
The trends with different shear-shear surveys reflect those already discussed in reference to Fig.~\ref{fig:sigma8}.
As a result, the joint constraint (right panel of Fig.~\ref{fig:s8_z_IA}) shows the halo model with $A=1$, $\beta=1.44$ (circle) is now slightly lower than when IAs are ignored.
This is consistent with the contours shown in Fig.~\ref{fig:7} which have a lower $\sigma_8$ at $A=1$ than $A=0$.

We next allow $A$ and $\beta$ 
 to vary within the halo model,
and we marginalise over both parameters with flat wide priors using only the shear-shear correlation data. The results are 
shown by the stars in Fig.~\ref{fig:s8_z_IA}.
As expected, the uncertainties are larger than when these parameters are kept fixed. The
best fit values are about the same as when no IAs are included. We would expect this from  Fig.~\ref{fig:7} in which the $A=0$ line cuts a cross-section through the banana-shaped degeneracy contour from shear-shear correlations alone which is roughly in the middle of the $\sigma_8$ range allowed by the whole contour.

Joint constraints from the combined cosmic shear and IA data (triangles) gives tighter constraints on $\sigma_{\rm 8}$
than from cosmic shear data alone, as expected. 
The constraint from all shear-shear surveys (right hand panel) is consistent with the contours in Fig.~\ref{fig:7}, where it is clear that the joint constraint (solid line) gives a very similar $\sigma_8$ value to the cosmic shear constraint without IAs (a slice through the cosmic shear contour at A=0).

The variation between different IA treatments are most strongly pronounced for the case of the VIRMOS-DESCART survey (highest median redshift of the four survey points in Fig.~\ref{fig:s8_z_IA}).
Although it has the highest median redshift, we have seen from Fig.~\ref{fig:n(z)} that it also contains
a large fraction of low-redshift galaxies and hence most physically close galaxies. This increases the II IA signal, that is significant for the range of scales probed by VIRMOS-DESCART. This boosts the amplitude of the predicted shear-shear correlation function points, requiring lower $\sigma_{\rm 8}$ to compensate. This behaviour produces the larger bias observed in VIRMOS-DESCART on going from no IA to the halo model.

The new $\sigma_8$ value can be higher or lower than when IAs are ignored (horizontal crosses), depending on the dataset in question. However all the new cosmology constraints (triangles) lie easily within the 68$\%$ confidence limits of the constraint without IA, including the constraint for all shear-shear surveys combined.

\begin{figure}
\center
\epsfig{file=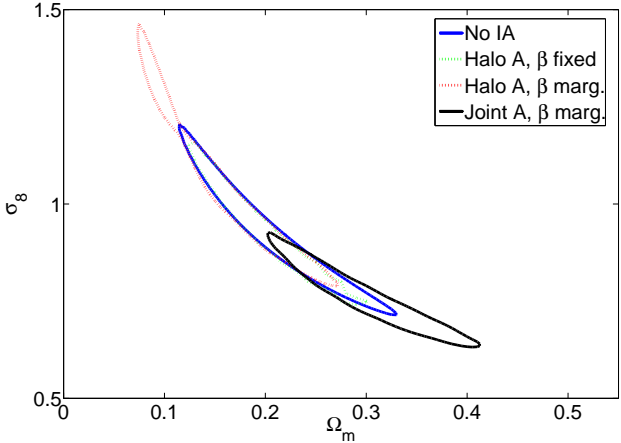,height=6cm,angle=0}
\caption{Combined 68$\%$ confidence limits in $\sigma_{\rm 8}$-$\Omega_{\rm m}$ parameter space for the 100 deg$^2$
shear-shear correlation function dataset. Constraints shown without IAs (blue solid contour); using the halo model with $A=1$, $\beta=1.44$ (green dotted contour); halo model, marginalised over $A$ and $\beta$ (red dotted contour); joint
shear-shear and shear-position constraints, marginalised over $A$ and $\beta$ (black solid contour).
All other cosmological parameters were fixed at their fiducial values.
}
\label{fig:Cls_s8_Odm_joint}
\end{figure}

We repeated the analysis for constraints in $\sigma_{\rm 8}$-$\Omega_{\rm m}$ space. All other cosmological parameters were held at their fiducial values. Fig.~\ref{fig:Cls_s8_Odm_joint} shows the 68$\%$ confidence contours for cosmic shear data without IAs (blue solid contour), that are the same as the dark solid lines in Fig.~\ref{fig:CLs_s8Odm}.
The halo model with fixed IA parameters (green dotted contour) shows similar behaviour to that discussed in Fig. \ref{fig:CLs_s8Odm}, favouring the high $\sigma_{\rm 8}$ and low $\Omega_{\rm m}$ region. Marginalising over the IA amplitude and luminosity-dependence parameters, $A$ and $\beta$ (red dotted contour), relaxes the constraint, extending the degeneracy further into the high $\sigma_{\rm 8}$, low $\Omega_{\rm m}$ region. The joint constraints (black solid contour) using both shear-shear and shear-position information show a preference for low $\sigma_{\rm 8}$ and high $\Omega_{\rm m}$ but are still consistent with our fiducial cosmology.
The preference for higher $\Omega_{\rm m}$ values occurs 
because, as noted in \citet{hirataea07}, the shear-position data favours slightly steeper correlation functions than predicted by a standard fiducial cosmology (see also the L6 panel of Fig.~\ref{fig:corrfns_hirata}). Increasing $\Omega_{\rm m}$ produces a correlation function that falls off faster at larger scales because the linear theory matter power spectrum peak is moved to the right (smaller scales) on increasing $\Omega_{\rm m}$ due to the earlier epoch of matter-radiation equality.

This new constraint on cosmology from cosmic shear self-consistently takes into account IA effects and we find
$\Omega_{\rm m}= 0.18 \pm 0.09$,
$\sigma_8= 0.88 \pm 0.18$ at 68 $\%$ 
 confidence, as opposed to the results in which IAs are ignored, that give
$\Omega_{\rm m}=0.20 \pm 0.10$,
$\sigma_8=0.86 \pm 0.21$.
When we constrain simultaneously with shear-shear and shear position data we find
$\Omega_{\rm m}= 0.29 \pm 0.09$, $\sigma_8= 0.74 \pm 0.13$,
after marginalising over IA parameters.
There is a strong degeneracy between $\sigma_8$ and $\Omega_{\rm m}$ when using current cosmic shear data alone, so it is better to look at the full two-dimensional contours (Fig.~\ref{fig:Cls_s8_Odm_joint}) or at cross-sections.
When we fix $\Omega_{\rm m}=0.3$ our best self-consistent
shear-shear 
result including IAs (right hand panel of Fig.~\ref{fig:s8_z_IA}) is $\sigma_8=0.75 \pm 0.04$ as compared to that when IAs are ignored of $\sigma_8=0.75 \pm 0.03$. A joint constraint from shear-shear and shear-position data, marginalising over IA parameters, gives
$\sigma_8=0.75 \pm 0.03$.

\section{Conclusions}
\label{sec:conclusions}

IAs are expected to be an important source of systematic error in cosmic shear measurements
if completely ignored. Several previous constraints on cosmology from shear-shear correlation functions have considered the possible levels of contamination from IAs, but the recent 
constraints presented have not included the impact of IAs~\citep{benjaminea07,fuea08_mnras,schrabbackea09}. These surveys are therefore referred to as ``cosmic shear" surveys.

In this paper we use three physically motivated IA models, of increasing complexity, to demonstrate
how constraints on the amplitude of fluctuations $\sigma_8$ and matter density $\Omega_{\rm m}$ are changed depending on the assumptions about IAs.
The size of the change in $\sigma_8$ depends on the redshift distribution of galaxies in the shear-shear survey and on the range of scales probed. In particular this affects the interplay between the two different types of IA (II and GI) which have competing effects on cosmological constraints.
A simple examination of IAs which assumes all galaxies are like the SDSS Main Red L4 sample suggests that $\sigma_8$ has previously been underestimated by about one 
standard deviation.
The underestimate appears to be larger using the non-linear matter power spectrum in the linear alignment model for IAs than when using the linear matter power spectrum.
On using the halo model for IAs the bias depends significantly on the range of scales probed by the survey due to the larger effect of II at intermediate scales which biases the measured $\sigma_8$ downwards when taken into account.

We have performed the first constraints on multiple IA model parameters from observed shear-position correlation functions using a physically motivated model for IAs. We also show the first simultaneous constraints on IA and cosmological parameters from either shear-position or shear-shear correlation functions. As expected, the majority of the constraint on cosmology comes from the shear-shear correlation functions, and the majority of the constraint on IAs comes from the shear-position correlation functions.

We have used a motivated model for luminosity and colour evolution of shear-shear galaxy samples as a function of redshift to improve our constraints on cosmology from shear-shear correlation data in the presence of the halo model for IAs. In general this reduces the overall effect of IAs on cosmology,
which 
we attribute to the decreased contribution from IA at low redshifts. We consider constraints on the amplitude of fluctutions $\sigma_8$ from shear-shear data alone with fixed IA amplitude and luminosity dependence parameters, and compare this with constraints after marginalising over these parameters.
As expected the constraints are weaker after marginalisation, but are not biased significantly. This is due to the relatively flat degeneracy between $\sigma_8$ and the IA amplitude arising from the competing effects of GI and II terms, and the ability of shear-shear information alone to place some constraint on the IA amplitude if other cosmological parameters are held fixed.

The model we used for the luminosity and colour evolution of the shear-shear galaxy sample could be improved in its complexity, for example by allowing a population of faint blue galaxies at high redshift. Furthermore we assumed the shear-shear galaxy sample had a single luminosity and colour at each redshift, whereas a more sophisticated analysis would integrate over ranges in each.

A joint analysis of shear-shear and shear-position correlation functions has been discussed as a promising tool for removing IAs from future cosmic shear
datasets~\citep{zhang08,bernstein_2008,joachimi_bridle_2009}.
We perform the first joint analysis of these correlation functions on observational data.
The uncertainties on the amplitude of fluctuations $\sigma_8$ are now reduced to the size found when IAs are ignored.
The estimated value of $\sigma_8$ is either decreased or increased relative to the case where IAs are ignored. However, for the combined constraints from the~\cite{benjaminea07} shear-shear data compliation the overall $\sigma_8$ value is barely changed on performing the full joint analysis of shear-shear and shear-position correlation functions with IAs, relative to the conventional analysis in which IAs are ignored and only shear-shear information is used.

The constraints in the two-dimensional parameter space of $\sigma_8$ and $\Omega_{\rm m}$ show a bigger change, that we have attributed to the slope of the shear-position correlation function preferring higher $\Omega_{\rm m}$ values. This raises the question of how to incorporate cosmological parameters into IA predictions. For example we could have considered the IA power spectra from our fiducial cosmological model as template functions which are allowed to vary only in amplitude. It also places a new burden of accuracy on measurements of the shear-position correlation function, that in turn depend on measurements of galaxy biasing for their interpretation.

We have not repeated the full analysis of \citet{benjaminea07}, who consider several other potential sources of systematic uncertainty including photometric redshift and shear calibration errors. We defer this to future investigations that could incorporate the very latest cosmic shear data.
The recent re-analysis of the COSMOS cosmic shear data~\citep{massey_growth2007_mnras} by \citet{schrabbackea09} investigates a number of potential cosmic shear systematics including IAs.
The photometric redshifts are used to produce the first cosmic shear tomographic cross-correlation functions which allow a much better control of IAs. Furthermore they investigate the effect of removing red galaxies from the cosmic shear sample which should help to reduce the effect of IAs further. This would have to be carefully modelled in the predictions of the IA contribution to the shear-shear correlation functions.

We believe that future cosmic shear analyses should include a model for IAs and use the existing galaxy survey data to help constrain this simultaneously with cosmology, as set out in this paper.
The modelling in this paper could be improved by extending the flexibility in the IA model, for example by allowing the one- and two-halo terms to vary independently in amplitude, and allowing an unknown amplitude for the IAs for blue galaxies. Furthermore the amplitude parameters could be free functions of redshift and even scale.
The IA models we have used specify simultaneously the two types of IA, II and GI. However, a more general model of IAs can lead to independent uncertainties in each of the source terms of these effects~\citep[see the bias and correlation terms of the intrinsic convergence with the lensing potential in]{bernstein_2008}.

However, if such a large degree of flexibility were allowed with current shear-position information then the constraining power of current cosmic shear data would be significantly reduced. Therefore it is important to obtain better measurements of the shear-position correlation functions, and/or improve models of IAs to reduce the number of free parameters.
In this analysis we have used measurements of the galaxy shear-position correlation function made from a spectroscopic galaxy redshift survey. 
It should be possible to make these measurements from the same data used for the cosmic shear survey itself, however this has not yet been done.
This will require a careful understanding of the survey mask as well as a measurement of the galaxy biasing to allow the shear-position correlation function to be related to the shear-density correlation function. 
In addition there are constraints in the literature from the shear-shear correlation functions of low redshift galaxy samples as well as for higher redshift spectroscopic galaxy samples which can be used to provide further constraints on the IA contributions.


\section*{Acknowledgements}

We are grateful to Catherine Heymans, Jonathan Benjamin, Katarina Markovic,
Benjamin Joachimi, 
Rachel Mandelbaum, Filipe Abdalla, Ofer Lahav, Henk Hoekstra, Lisa Voigt Alexandre Refregier, Dugan Witherick and Shaun Thomas for helpful discussions.
We thank CEA Saclay for hospitality where part of this work was carried out.
SLB acknowledges support from a Royal Society University Research Fellowship.

\bsp

\bibliographystyle{mn2e}

\appendix

\section{Models for mean luminosity and red galaxy fraction}
\label{sec:clfmodel}
To match the luminosity and colour dependence of the IA model in
the cosmic shear data 
we use a model for the conditional luminosity function (CLF) closely following~\citet{cooray2006}.  The conditional
luminosity function $\Phi(L|M,z)$ gives the mean number of galaxies in a dark matter halo of mass $M$ at redshift
$z$ with luminosity in the range $[L,L+dL]$.  This function provides a way to add luminosity dependence to the halo
model description of galaxy statistics.

Following the usual practice in the halo model, the CLF model is constructed from separate models for the central and
satellite galaxies.  The first step is to specify the mean luminosity of the central galaxy in a halo of mass
$M$, 
$L_c(M,z)$,
which we model as a broken power-law in mass according to
Eq.~3 
of \citet{cooray2006}; where the mean luminosity of the
central galaxy increases quickly with mass for low halo masses and then flattens off for larger halo masses.  The total
luminosity of a halo should include contributions from satellite galaxies for masses greater than some threshold, $\msat$,
\begin{equation}\label{eq:Ltot}
  \ltot(M,z) =
  \begin{cases}
    L_c(M,z) & M\le\msat(z)\\
    L_c(M,z)\,\left(\frac{M}{\msat(z)}\right)^{\beta_s(z)} & M > \msat(z)
  \end{cases}
\end{equation}
where $\beta_s=0.8$, $\msat(z) \equiv \msat\, \text{erf}(5\,z)$ and $\msat=10^{13}\,h^{-1}M_{\odot}$.  This particular $z$ dependence
is an ad-hoc adjustment to allow lower mass halos to host satellites at low $z$.

In the simplest model the CLF for central galaxies is just a delta-function centered at $L_c$, although \citet{cooray2006}
suggests a ``rounded'' step to allow for scatter in the minimum mass halo hosting a galaxy,
\begin{equation}
  \phicen(L|M,z) = \Theta_{H}\left(M-\mcen\right)\,\text{Log-Normal}(L|L_c,\sigma_c),
\end{equation}
where $\sigma_c=1.15$, $\Theta_H$ is the Heaviside step function and $\mcen=10^{10}\, h^{-1}M_{\odot}$.

The satellite CLF is modeled as a power-law in luminosity with maximum luminosity equal to half the central galaxy
luminosity (so that satellites do not dominate the total luminosity of the halo),
\begin{equation}
  \phisat(L|M,z) = \Theta_{H}\left(\frac{L_c}{2}-L\right)\,
  A(M,z)\, \frac{1}{L}.
\end{equation}
The amplitude, $A(M,z)$, is determined by requiring $\int_{\lmin}^{\lmax} dL L \phisat(L|M,z) = \ltot-L_c$, giving,
\begin{equation}\label{eq:phisatamp}
  A(M,z) \approx
  \frac{\ltot(M,z) - L_c(M,z)}{\text{min}\left(\lmax,L_c(M,z)/2\right) - \lmin(z) }
  . 
\end{equation}
For modeling the luminosity dependence in the 100~sq.~deg. lensing data sets, we set the integration limits $\lmin, \lmax$
to match the apparent magnitude limits given in table~1 of \citet{benjaminea07}.
Note the luminosity integration limits become redshift dependent when matching fixed apparent magnitude limits.
For the upper integration limit we further set
$\lmax(M,z)={\rm min}\left(L_c(M,z)/2, L_{\rm max, survey}(z)\right)$.

The HOD functions giving the number of central and satellite galaxies in a halo of mass $M$ are obtained by integrating the CLFs
over luminosity,
\begin{align}
  N_{g}^{\rm cen,sat}(M,z) &= \int_{\lmin}^{\lmax} dL\, \Phi^{\rm cen,sat}(L|M,z)
  . 
\end{align}

Integrating over halo mass, we can relate the CLFs to the redshift distribution of the galaxy sample as,
\begin{align}
  \frac{dN}{dzd\Omega}(z) &= \chi^2(z)\,\frac{d\chi}{dz}(z)\,
  \int dM\, n(M,z)
  \notag\\
  &\times\left(\ngc(M,z)+\ngs(M,z)\right),
\end{align}
where $\chi(z)$ is the comoving distance as a function of redshift\footnote{assuming zero spatial curvature.}.
We match the redshift distributions from \citet{benjaminea07} with the CLF prediction to fit the (redshift-dependent)
normalization of the central galaxy luminosity $L_c(M,z)$.

\begin{figure}
\centering
\includegraphics[scale=0.5]{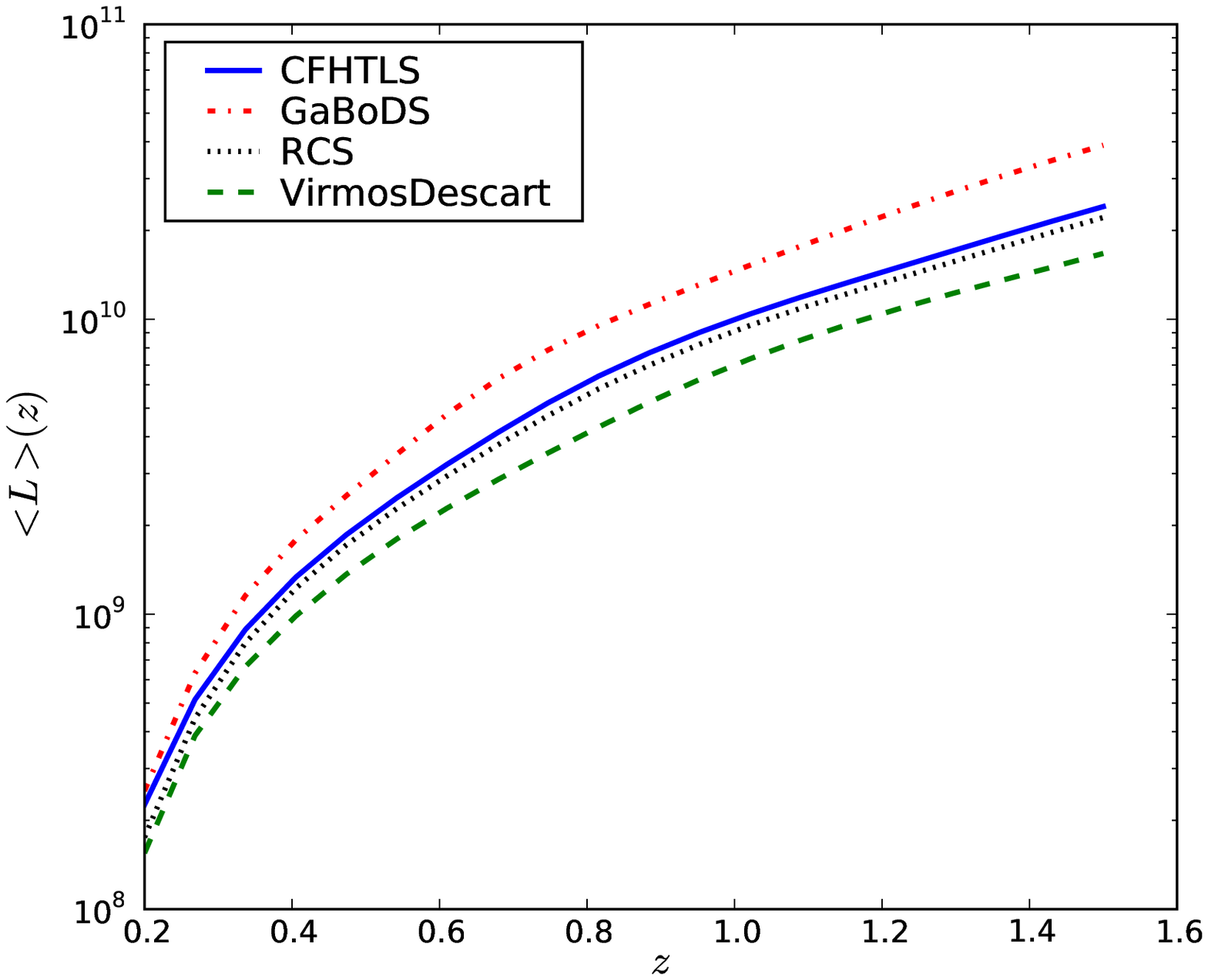}
\caption{\label{fig:lbar}Mean luminosity model versus redshift for each of the 4 lensing surveys.}
\end{figure}

With the parameters of our CLF model matched to the magnitude limits and redshift distribution of each lensing survey,
we calculate the mean luminosity of each galaxy sample as,
\begin{equation}
  \left<L\right>(z) = \frac{\int dM\, n(M,z) \int dL\, L\,\Phi(L|M,z)}
  {\int dM\, n(M,z)\int dL\, \Phi(L|M,z)}
 . 
\end{equation}
We show our model for $\left<L\right>(z)$ for each of the
four 
lensing surveys in
 Fig.
 \ref{fig:lbar}.

We again follow \citet{cooray2006} to model the fraction of red central galaxies as a function of halo mass as a ``rounded''
step function,
\begin{equation}
  \fredcen(M,z) = \fredcen(z)\half \left(1+\text{erf}\left(\frac{\ln(M)-\ln(M_{\text{cen}})}{\sigma_{\text{red,cen}}}\right) \right)
  . 
\end{equation}
This approaches the constant value $\fredcen(z)$ as $M$ becomes large and goes to zero for small $M$.  Note this has no
luminosity dependence.

The fraction of red satellites in a halo of mass $M$ is modeled as,
\begin{equation}
  \fredsat(M,L,z) = g_{\text{red,sat}}(z)\left(g(M,z) + h(L,z)\right) + \fredsat(z)
\end{equation}
where $g(M,z)$ and $h(L,z)$ are ``rounded step functions'' in mass and luminosity, respectively.  For low masses and
luminosities, the red satellite fraction is the constant $\fredsat(z)$ while for high masses and luminosities it is
the different constant $2g_{\text{red,sat}}(z)+\fredsat(z)$.
For simplicity, we set $\fredcen$, $\fredsat$, and $g_{\text{red,sat}}$ to the $z$-independent constants 0.6, 0.4, and 0.2.

From these pieces, the fraction of all red galaxies as a function of redshift is,
\begin{align}
  f_{\text{red}}(z) &= \frac{1}{N_g(z)}\int dM\, n(M,z)
  \int dL\,
  \notag\\
  &\times[\phicen(L|M,z)\fredcen(M,z)
  \notag\\
  &\qquad+\phisat(L|M,z)\fredsat(M,L,z)] ,
\end{align}
where $N_g(z)$ is the comoving number density of galaxies in the sample.

\begin{figure}
\centering
\includegraphics[scale=0.36]{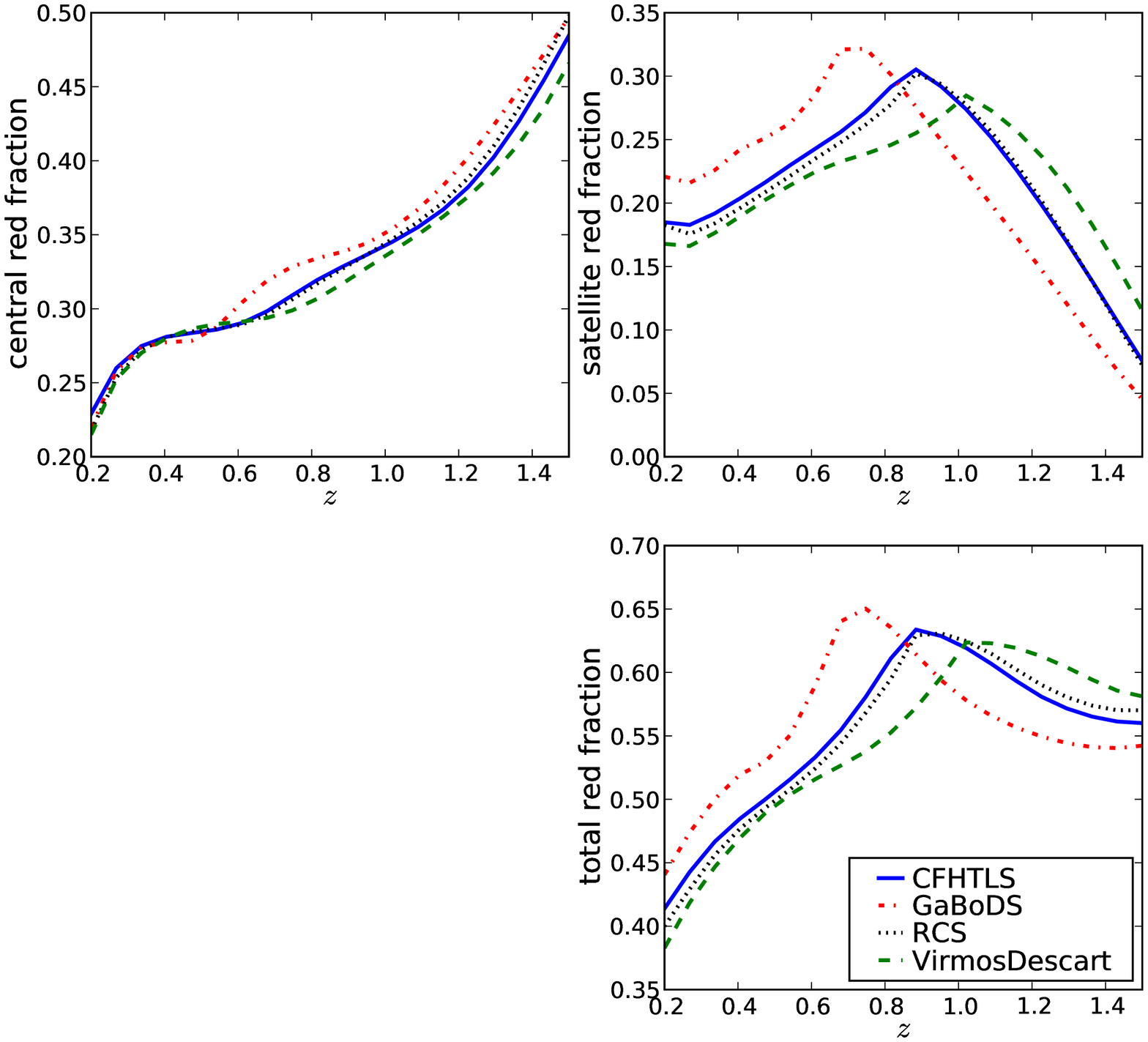}
\caption{\label{fig:fred}Red galaxy fraction versus redshift for each of the 4 lensing surveys.  The top panels show the red fraction
for central and satellites separately in our halo model while the bottom panel shows the red fraction for all galaxies.}
\end{figure}

Our models for $f_{\rm red}(z)$ for each of the 4 lensing surveys are shown in fig.~\ref{fig:fred}.
Ignoring redshift evolution in $f_{\rm red,cen}$ and $f_{\rm red,sat}$ is almost certainly incorrect and, in combination with the
fixed magnitude cutoffs in our model, tends to select
excess red galaxies at high redshift (when normalized to give reasonable red fractions at low redshift).  Because the IA halo model
includes contributions only from red galaxies, we consider this model for $f_{\rm red}$ to give an upper bound on the possible IA
contamination to the weak lensing data.  This is suitable for the goal in this paper of assessing the impact of IAs as systematic
errors in lensing surveys.  To improve the predictive power of our IA halo model in the future it will be important to better understand
the redshift evolution of the red fraction of galaxies.

\label{lastpage}

\end{document}